\documentstyle[12pt]{article}

\begin{document} 

\title{Faster-than-light effects and negative group delays in
optics and electronics, and their applications}

\author{Raymond Y. Chiao,  Jandir M. Hickmann\thanks{%
$\;$ $\;$On leave from the Department of Physics, Universidade Federal de Alagoas,
Brazil; supported by the Brazilian agency CNPq.} $\;$ and Daniel Solli \\
{\it Department of Physics, University of California} \\
{\it Berkeley, CA, 94720-7300, USA}}

\date{}

\maketitle 

\begin{abstract}
Recent manifestations of apparently faster-than-light effects confirmed our
predictions that the group velocity in transparent optical media can exceed
c. Special relativity is not violated by these phenomena. Moreover, in the
electronic domain, the causality principle does not forbid negative group
delays of analytic signals in electronic circuits, in which the peak of an
output pulse leaves the exit port of a circuit \textit{before} the peak of
the input pulse enters the input port. Furthermore, pulse distortion for
these ``superluminal'' analytic signals can be negligible in both the
optical and electronic domains. Here we suggest an extension of these ideas
to the microelectronic domain. The underlying principle is that negative
feedback can be used to produce negative group delays. Such negative group
delays can be used to cancel out the positive group delays due to
``transistor latency'' (e.g., the finite RC rise time of MOSFETS caused by
their intrinsic gate capacitance), as well as the ``propagation delays'' due
to the interconnects between transistors. Using this principle, it is
possible to speed up computer systems.\end{abstract}

\section{INTRODUCTION}
Recent optical experiments at Princeton NEC \cite{wang} have verified the
prediction by the one of the authors and his co-workers that superluminal
pulse propagation can occur in transparent media with optical gain \cite%
{chiao93}. \ These experiments have shown that a laser pulse can propagate
with little distortion in an optically pumped cesium vapor cell with a group
velocity greatly exceeding the vacuum speed of light $c$.\ In fact, the
group velocity for the laser pulse in this experiment was observed to be $%
negative$: The peak of the output laser pulse left the output face of the
cell $before$ the peak of the input\ laser pulse entered the input face of
the cell. \ This pulse sequence is counter-intuitive.

We also performed some earlier experiments on the speed of the quantum
tunneling process \cite{steinberg}. \ We found that a photon tunneled
through a barrier at an effective group velocity which was faster than $c$,
i.e., at a ``superluminal'' velocity. \ In these experiments, spontaneous
parametric down-conversion was used as a light source which emitted
randomly, but simultaneously, two photons at a time, i.e., photon ``twins.''
\ These photons were detected by means of two Geiger counters (silicon
avalanche photodiodes), so that the time at which a ``click'' was registered
is interpreted as the time of arrival of the photon. \ Coincidence detection
was used to detect these photon twins. \ One photon twin traverses a tunnel
barrier, whilst the other traverses an equal distance in the vacuum.

The idea of the experiment was to measure the time of arrival of the
tunneling photon with respect to its twin, by measuring the time difference
between the two ``clicks'' of their respective Geiger counters. \ (We
employed a two-photon interference effect in order to achieve sufficient
time resolution.) \ The net result was surprising: On the average, the
Geiger counter registering the arrival of the photon which tunneled through
the barrier clicked $earlier$ than the Geiger counter registering the
arrival of the photon which traversed the vacuum.\ \ This indicates that the
process of tunneling in quantum physics is superluminal.

The earliest experiment to demonstrate the existence of faster-than-$c$
group velocities was performed by Chu and Wong at Bell Labs. \ They showed
that picosecond laser pulses propagated superluminally through an absorbing
medium in the region of anomalous dispersion inside the optical absorption
line \cite{chu}. \ This experiment was reproduced in the millimeter range of
the electromagnetic spectrum by Segard and Macke \cite{segard}. \ These
experiments verified the prediction of Garrett and McCumber \cite{garrett}
that Gaussian-shaped pulses of electromagnetic radiation could propagate
with faster-than-$c$ group velocities in regions of anomalous dispersion
associated with an absorption line. \ Negative group velocities were also
observed to occur in these early experiments.\ 

These counter-intuitive pulse sequences were also seen\ to occur in
experiments on electronic circuits \cite{morgan}. \ In the first of these
experiments, we used an electronic circuit which consisted of an operational
amplifier with a negative feedback circuit containing a passive RLC network.
\ This circuit produced a negative group delay similar to that observed in
the optical experiment performed at Princeton NEC: The peak of the output
voltage pulse left the output port of the circuit $before$ the peak of the
input\ voltage pulse entered the input port of the circuit. Such\ a
seemingly anti-causal phenomenon does not\ in fact violate the principle of
causality, since there is sufficient information in the early portion of any
analytic voltage waveform to reproduce the entire waveform earlier in time.
We showed that causality is solely connected with the occurrence of
discontinuities, such as ``fronts'' and ``backs'' of signals, and not with
the peaks in the voltage waveform, and, therefore, that causal loop
paradoxes could never arise \cite{john}.

We propose to apply these counter-intuitive ideas to the design of
microelectronic devices \cite{sedra}. \ This is timely, since it is widely
believed that Moore's law for microprocessor performance will fail to hold
in the next decade due to a ``brick wall'' arising from fundamental physical
limitations \cite{packan}. Therefore, there have been many proposals for new
transistor technologies to try to solve this problem \cite{taur}\cite%
{geppert}. \ At the present time, the ``transistor latency'' problem is one
of the main factors limiting computer performance, although the
``propagation delays'' due to the RC time constants in the interconnects
between individual transistors on a computer chip are beginning to be
another serious limiting factor. \ As the scale of microprocessor circuits
fabricated on a silicon wafer is reduced to become ever smaller in size, the
transistor switching time becomes increasingly faster, but the propagation
delay from transistor to neighboring transistor becomes increasingly longer %
\cite{miller}. \ This will still be true even after new technologies to
replace MOSFETS with faster devices is implemented. \ \ 

The propagation delays of interconnects arise from a combination of the
resistivity of the evaporated metal wire connecting two nearby transistors,
and the dielectric constant of the insulator which supports the
interconnecting wire. \ The solution to this problem being currently
implemented in industry is to use copper interconnects instead of aluminum
(which has traditionally been used). \ Another solution is to use insulators
with a lower dielectric constant to support the interconnecting wires. \
These solutions reduce propagation delays, but do not eliminate them
altogether.

Here we suggest a radically different approach which in principle can
eliminate all kinds of delays by implementing the concept of negative group
delay in conjunction with negative feedback. \ For example, this will allow
us to eliminate the positive propagation delay from an interconnect by
exactly compensating for this delay with an equal, but opposite, negative
group delay. \ We therefore suggest that the architecture of microprocessors
should be changed to incorporate negative feedback elements between logic
gates. \ 

The compensation of propagation delays, if implemented properly, would also
lead to a path independence for the routing time of logic pulses throughout
the computer system. \ This leads to a novel solution of the ``clock skew''
or ``clock synchronization'' problem, in which there arises a skewed
distribution of arrival times of logic pulses at a final logic gate, due to
the fact that these pulses are routed through different paths on a computer
chip. \ The phenomenon of clock skew prevents the use of higher clock rates,
because the current solution to this problem is to deliberately add an extra
delay to an early-arriving pulse, such that it arrives simultaneously with a
late-arriving pulse at the final gate \cite{miller}. Thus, existing computer
clock speeds are determined in practice by the propagation delay of the $%
longest$ routing path. \ 

We believe that the clock skew problem can be eliminated by eliminating
propagation delays altogether. \ Since there would no longer\ be any
appreciable delays for a pulse to propagate from logic gate to logic gate,
the routing time of a logic pulse to a final logic gate could become largely
independent of the path taken by this pulse inside the computer. \ Much
faster computer systems should result.

\section{GENERAL PRINCIPLES FOR GENERATING NEGATIVE GROUP DELAYS}

\subsection{Negative group delays necessitated by the golden rule for
operational amplifier circuits with negative feedback}

In Figure 1, we show an operational amplifier with a signal entering the
noninverting (+) port of the amplifier. \ The output port of the amplifier
is connected\ back to the inverting ($-$) port of the amplifier by means of
a black box, which represents a passive linear circuit with an arbitrary
complex transfer function $\widetilde{F}\left( \omega \right) $ for a signal
at frequency $\omega $. \ We thus have a linear amplifier circuit with a
negative feedback loop containing a passive filter. \ In general, the
transfer function of any passive linear circuit, such as a RC low-pass
filter, will always lead to a $positive$ propagation delay through the
circuit.  

However, for operational amplifiers with a sufficiently high gain-feedback
product, the voltage difference between the two input signals arriving at
the inverting and noninverting inputs of the amplifier must remain small at
all times. \ The operational amplifier must therefore supply a signal with a 
$negative$ group delay at its output, such that the $positive$ delay from
the passive filter is exactly canceled out by this negative delay at the
inverting ($-$) input port. \ The signal at the inverting ($-$) input port
will then be nearly identical to that at the noninverting (+) port, thus
satisfying the golden rule for the voltage difference at all times. \ The
net result is that this negative feedback circuit will produce an output
pulse whose peak leaves the output port of the circuit $before$ the peak of
the input pulse arrives at the input port of this circuit.

In Figure 2, we show experimental evidence for this counter-intuitive
behavior for the special case of an RLC tuned bandpass circuit in the
negative feedback loop \cite{morgan}. \ The peak of an output pulse is $%
advanced$ by 12.1 milliseconds relative to the input pulse. \ The output
pulse has obviously not been significantly distorted with respect to the
input pulse by this linear circuit. \ Also, note that the size of the
advance of the output pulse is comparable in magnitude to the width of the
input pulse. \

That causality is not violated is demonstrated in a second experiment, in
which the input signal voltage is very suddenly shorted to zero the moment
it reaches its maximum. \ The result is shown in Figure 3. \ By inspection,
we see that the output signal is also very suddenly reduced to zero voltage
at essentially the same instant in time that the input signal has been
shorted to zero. \ This demonstrates that the circuit cannot advance in time
truly $discontinuous$ changes in voltages, the only points on the signal
waveform which are connected by causality \cite{john}. \ However, for the $%
analytic$ changes of the input signal waveform, such as those in the early
part of the Gaussian input pulse which we used, the circuit\ evidently has
the ability to extrapolate the input waveform into the future, in such a way
as to reproduce the output Gaussian pulse peak $before$ the input pulse peak
has arrived. \ In this sense, the circuit $anticipates$ the arrival of the
Gaussian pulse.

\subsection{The golden rule and the inversion of the transfer function of
any passive linear circuit}

Now we shall analyze under what conditions the golden rule holds and
negative group delays are produced. \ In Figure 1, $\widetilde{A}\left(
\omega \right) $ denotes the complex amplitude of an input signal of
frequency $\omega $ into the noninverting (+) port and $\widetilde{B}\left(
\omega \right) $ refers to that of the feedback signal into the inverting ($-
$) port of the amplifier. \ The output signal $\widetilde{C}\left( \omega
\right) $ is then related to the feedback signal $\widetilde{B}\left( \omega
\right) $ by means of the complex linear feedback transfer function $%
\widetilde{F}\left( \omega \right) $ (the black box) as follows:%
\begin{equation}
\widetilde{B}\left( \omega \right) =\widetilde{F}\left( \omega \right) 
\widetilde{C}\left( \omega \right) .
\end{equation}%
The voltage gain of the operational amplifier is characterized by the active
complex linear transfer function $\widetilde{G}\left( \omega \right) $,
which amplifies the difference of the voltage signals at the (+) and ($-$)
inputs to produce an output signal as follows:%
\begin{equation}
\widetilde{C}\left( \omega \right) =\widetilde{G}\left( \omega \right)
\left( \widetilde{A}\left( \omega \right) -\widetilde{B}\left( \omega
\right) \right) .
\end{equation}%
Defining the total complex transfer function $\widetilde{T}\left( \omega
\right) \equiv \widetilde{C}\left( \omega \right) /\widetilde{A}\left(
\omega \right) $ as the ratio of the output signal $\widetilde{C}\left(
\omega \right) $ to input signal $\widetilde{A}\left( \omega \right) $, we
obtain for the total transfer function,%
\begin{equation}
\widetilde{T}\left( \omega \right) =\frac{\widetilde{G}\left( \omega \right) 
}{1+\widetilde{F}\left( \omega \right) \widetilde{G}\left( \omega \right) }.
\end{equation}%
If the gain-feedback product is very large compared to unity, i.e., 
\begin{equation}
\left| \widetilde{F}\left( \omega \right) \widetilde{G}\left( \omega \right)
\right| >>1,
\end{equation}%
we see that to a good approximation this leads to the inversion of the
transfer function of any passive linear circuit by the negative feedback
circuit, i.e.,%
\begin{equation}
\widetilde{T}\left( \omega \right) \approx 1/\widetilde{F}\left( \omega
\right) =\left( \widetilde{F}\left( \omega \right) \right) ^{-1}.
\end{equation}%
This also implies through Eq. (2), that the golden rule,%
\begin{equation}
\widetilde{A}\left( \omega \right) \approx \widetilde{B}\left( \omega
\right) ,
\end{equation}%
holds under these $same$ conditions. \ Equation (5) also implies that the
negative feedback circuit shown in Figure 1 can completely undo any
deleterious effects, such as propagation delays, produced by a linear
passive circuit (whose transfer function is identical to $\widetilde{F}%
\left( \omega \right) $) when it is placed before this active device. \ 

In Figure 4 we show one example, where an RC low-pass filter is placed
before the negative feedback circuit. \ The positive propagation delay $\tau
_{\widetilde{F}\left( \omega \right) }$ due to this RC low-pass circuit, can
in principle be completely canceled out by the negative group delay produced
by the active circuit with the same RC circuit in its feedback loop. \ This
will be true in general for any linear passive circuit, if an identical copy
of the circuit is placed inside the negative feedback loop of the active
device. The group delay of the negative feedback circuit in the high
gain-feedback limit is then be given by 
\begin{equation}
\tau _{\widetilde{T}\left( \omega \right) }=\frac{d\arg \widetilde{T}\left(
\omega \right) }{d\omega }\approx \frac{d\arg \left( 1/\widetilde{F}\left(
\omega \right) \right) }{d\omega }=-\frac{d\arg \widetilde{F}\left( \omega
\right) }{d\omega }=-\tau _{\widetilde{F}\left( \omega \right) }.
\end{equation}%
This shows that the positive group delay from any linear passive circuit can
in principle be completely canceled out by the negative group delay from a
negative feedback circuit.

It is important to note that this negative feedback scheme places a
requirement on the gain-bandwidth product of the amplifier. For this active
circuit to advance the waveform, it must have a large gain at all of the
frequency components present in the signal. \ In particular, if we want to
counteract a particular $RC$ time delay, the amplifier must have a large
gain at frequencies greater than $1/RC$.

\subsection{Kramers-Kronig relations necessitate superluminal group
velocities, and Bode relations necessitate negative group delays}

These counter-intuitive results also follow quite generally from the
Kramers-Kronig relations in the optical domain \cite{landau}, and the Bode
relations in the electronic domain \cite{bode}. \ \ In the optical domain,
we have proved two theorems starting from the principle of causality, along
with the additional assumption of linearity, that superluminal group
velocities in any medium must generally exist in some spectral region, and
that for an amplifying medium, this spectral region must exist away from the
regions with gain, i.e., in the transparent regions outside of the gain
lines \cite{bolda}. Negative group delays in the electronic domain similarly
follow generally from the Bode relations. \ Thus, causality itself $%
necessitates$ the existence of these counter-intuitive phenomena.

\subsection{Energy transport\ by pulses in the optical and electronic domains%
}

In the optical domain, there has been a debate concerning whether or not the
velocity of energy transport by the wave packet can exceed $c$ when the
group velocity of a wave packet exceeds $c$. \ In the case of anomalous
dispersion inside an absorption line, Sommerfeld and Brillouin showed that
the energy velocity defined as,

\begin{equation}
v_{energy}\equiv \frac{\left\langle S\right\rangle }{\left\langle
u\right\rangle },
\end{equation}%
where $\left\langle S\right\rangle $ is the time-averaged Poynting vector
and $\left\langle u\right\rangle $ is the time-averaged energy density of
the electromagnetic wave, is $different$ from the group velocity \cite%
{brillouin}\cite{loudon}. \ Whereas the group velocity in the region of
absorptive anomalous dispersion exceeds $c$, they found that the energy
velocity is less than $c$. \ Experiments on picosecond laser pulse
propagation in absorptive anomalous dispersive media, however, show that
these\ laser pulses travel with a superluminal group velocity, and not with
the subluminal energy velocity of Sommerfeld and Brillouin \cite{chu}. \
Hence the physical meaning of this energy velocity is unclear. \ 

When the optical medium possesses gain, as in the case of laser-like medium
with inverted atomic populations, the question arises as to whether or not
to include the energy stored in the inverted atoms in the definition of $%
\left\langle u\right\rangle $ \cite{schulz-dubois}\cite{oughstun}. \ In
regions of anomalous dispersion outside of the gain line, and, in
particular, in a spectral region where the group-velocity dispersion
vanishes, a straightforward application of the Sommerfeld and Brillouin
definition of the energy velocity would imply that the group and energy
velocities both exceed $c$. \ The equality of these two kinds of wave
velocities arises because the pulses of light are propagating inside a
transparent medium with little dispersion. \ In particular, in the case when
the energy velocity is negative, the maximum in the pulse of energy leaves
the exit face of the optical sample $before$ the maximum in the pulse of
energy enters the entrance face, just like in the case of negative group
velocities.

In the case of the electronic circuit with negative feedback which produces
negative group delays, the question of when the peak of the energy arrives,
can be answered by terminating the output port of Figure 1 by a load
resistor, which connects the output to ground. \ The load resistor (not
shown) will be heated up by the energy in the \textit{output} pulse. \ It is
obvious that the load resistor will then experience the maximum amount of
heating when the peak of the Gaussian output pulse arrives at this resistor,
and that this happens when the peak of the output voltage waveform arrives.
\ For negative group delays, the load resistor will then heat up earlier
than expected. \ There is no mystery here: The operational amplifier can
supply the necessary energy to heat up the load resistor ahead of time. \
Hence the negative group and the negative energy delays are identical in
this case.

\subsection{Preliminary data demonstrating the elimination of propagation
delays from RC time constants}

In a recent experiment with the circuit shown in Figure 4, we obtained the
data in shown Figure 5 of the outputs from a square wave input into an RC
low-pass circuit, with (in the upper trace), and without (in the lower
trace) the negative feedback circuit inserted after it. \ It is clear by
inspection of the data in Figure 5 that the propagation delays due to the RC
time constant on both the rising and falling edges of the square wave have
been almost completely eliminated by the negative feedback circuit. \
However, there is a ringing or overshoot phenomenon accompanying the
restoration of the rising and falling edges. \ Since the CMOS switching
levels between logic states occur within 10\% of zero volts for LO signals,
and within 90\% of volt-level HI signals \cite{sedra}, the observed ringing
or overshoot phenomenon is not deleterious for the purposes of computer
speedup. \ 

It is clear from these data that not only the RC time constants associated
with transistor gates (the ``latency'' problem), but also the RC propagation
delays from the wire interconnects between transistors on a computer chip,
can in principle be eliminated\ by means of the insertion of negative
feedback elements.\ In particular, the finite rise time of a MOSFET arising
from its intrinsic gate capacitance can be eliminated. \ 

\section{CONCLUSIONS}

There is a widespread view among electrical engineers and physicists that
although the phase velocity can exceed the vacuum speed of light, the group
velocity can never do so. \ Otherwise, signals would be able to propagate
faster than light, since conventional wisdom equates the group velocity with
the signal velocity. \ Several generations of students have been taught
this. \ Many of the standard textbooks also teach this, but with some
qualifications, which unfortunately are\ not strong enough, so that the net
result is still quite misleading. \ For example, Born and Wolf in \textit{%
Principles of Optics} in their discussion concerning the group velocity
state the following \cite{born}:

\begin{quote}
If the medium is not strongly dispersive, a wave group will travel a
considerable distance without appreciable ``diffusion'' [i.e., dispersion].
In such circumstances, the group velocity, which may be considered as the
velocity of propagation of the group as a whole, will also represent the
velocity at which the energy is propagated. \ This, however, is not true in
general. In particular, in regions of anomalous dispersion the group
velocity may exceed the velocity of light or become negative, and \textit{in
such cases it no longer has any appreciable physical significance}.
[Emphasis added]\ 
\end{quote}

This statement is misleading. \ As a result, we have been blinded by our
misconceptions, and thereby been prevented from exploring and discovering
many new, interesting, and possibly important, phenomena, which should have
been discovered long ago. \ Some of these are only now being uncovered, and
some of these phenomena may in fact lead to important applications, such as
the speedup of computers.

\section*{ACKNOWLEDGMENTS}       
 
We thank Mohammad Mohajedi for helpful discussions. \ Some of this work was
supported by the ONR and by NASA.

  \end{document}